\documentclass[aps,prb,twocolumn,showpacs,preprintnumbers,amsmath,amssymb,superscriptaddress]{revtex4-2}

\usepackage[colorlinks=true,allcolors=blue]{hyperref}
\usepackage{xcolor}
\usepackage{txfonts}
\usepackage{graphicx}
\usepackage{color}
\usepackage{dcolumn}
\usepackage{bm}
\usepackage{amsmath}
\usepackage{feynmp}
\usepackage{float}
\usepackage{braket}
\usepackage{cleveref}

\newcommand{\nc}{\newcommand}
\nc{\bea}{\begin{eqnarray}}
\nc{\eea}{\end{eqnarray}}
\nc{\vx}{{\mathbf{r}}}
\nc{\vm}{\mathbf{m}}

\begin{document}

\title{Topology-constrained spin-wave modes of asymmetric antibimerons and their clusters}

\author{Pavel~A.~Vorobyev}
\email{p.vorobyev@unsw.edu.au}
\author{Daichi~Kurebayashi}
\author{Oleg~A.~Tretiakov}
\email{o.tretiakov@unsw.edu.au}
\affiliation{School of Physics, The University of New South Wales, Sydney 2052, Australia}

\date{April 10, 2026}

\begin{abstract}

Collective modes are a defining signature of coupled degrees of freedom, forming a bridge between understanding of interactions in condensed-matter systems and emergent functionality. Topological magnetic textures provide a natural platform to realize and control such collective modes at the nanoscale. Here we theoretically identify and characterize low-energy collective spin-wave excitations of isolated asymmetric antibimerons and their clusters in ultrathin ferromagnetic films. We demonstrate that an isolated asymmetric antibimeron supports a discrete spectrum of localized modes, reflecting its internal degrees of freedom. When multiple asymmetric antibimerons form a cluster, inter-texture coupling leads to the splitting of these modes into $N$-fold multiplets, where $N$ denotes the number of asymmetric antibimerons. To rationalize these findings, we introduce an effective coupled-oscillator model based on meron pairs that captures the essential collective dynamics of the system. This emergent classical mechanics description reveals that the motion of asymmetric antibimeron clusters can be understood in terms of well-defined normal modes governed by topology-constrained particle-like degrees of freedom. These results establish coupled asymmetric antibimerons as a tunable platform for spin-wave based nano-oscillators, whose normal-mode spectrum is controllable through cluster size, thus providing a programmable set of low-lying resonances for these nano-oscillators.

\end{abstract}

\maketitle



\setcounter{footnote}{0}
\renewcommand{\thefootnote}{\arabic{footnote}}

Collective modes in solids arise from the interplay of coupled degrees of freedom~\cite{Goldstein1980}. In magnetic systems, these excitations in the form of spin waves are especially attractive because they can carry information and enable high-frequency signal processing on nanometre scales~\cite{Chumak2015, Flebus2024}. Topological spin textures~\cite{Kosevich1990, Fert2017, GobelPhysRep2021, Zhou2025} offer a particularly rich setting for collective spin dynamics: their nontrivial topology and mutual interactions can generate localized collective modes with no analogue in collinear magnetic states. Understanding the origin of these modes, how they hybridize, and how they can be controlled is therefore essential both for the fundamental dynamics of topological matter and for future information-processing technologies~\cite{Parkin2008, Fert2013, Grollier2020, Pirro2021, Chumak2022, Sun2023, Everschor-Sitte2024}.

Magnetic skyrmions~\cite{Skyrme1962, Belavin1975, Muhlbauer2009, Robler2006, Nagaosa2013} have become the most prominent example of topological spin textures, and their generation~\cite{Koshibae2014, SoongGeun2018, Ohara2022}, detection~\cite{Neubauer2009, Heinze2011, Hirschberger2019}, and manipulation~\cite{Yu2012, Zhang2016, Barker2016, Amin2023} are now well established across a wide range of materials~\cite{Muhlbauer2009, Yu2010, Seki2012, Tokunaga2015, Legrand2020, Tokura2020, Meisenheimer2023}. For rotationally symmetric spin textures, such as circular skyrmions, this has led to a clear classification of low-energy eigenmodes~\cite{Petrova2011, Mochizuki2012, Schwarze2015, Mruczkiewicz2016, Kravchuk2019, Seki2020}, including rotational and breathing-like excitations. Yet for intrinsically asymmetric topological spin textures~\cite{Shen2020_2, FiM_Shen2022, Yu2024, Vorobyev2024}, where rotational symmetry is broken already at the level of an individual object, a general framework for low-energy mode classification is still lacking.

\begin{figure}[tbp]
	\centering
	\includegraphics[width=1\linewidth]{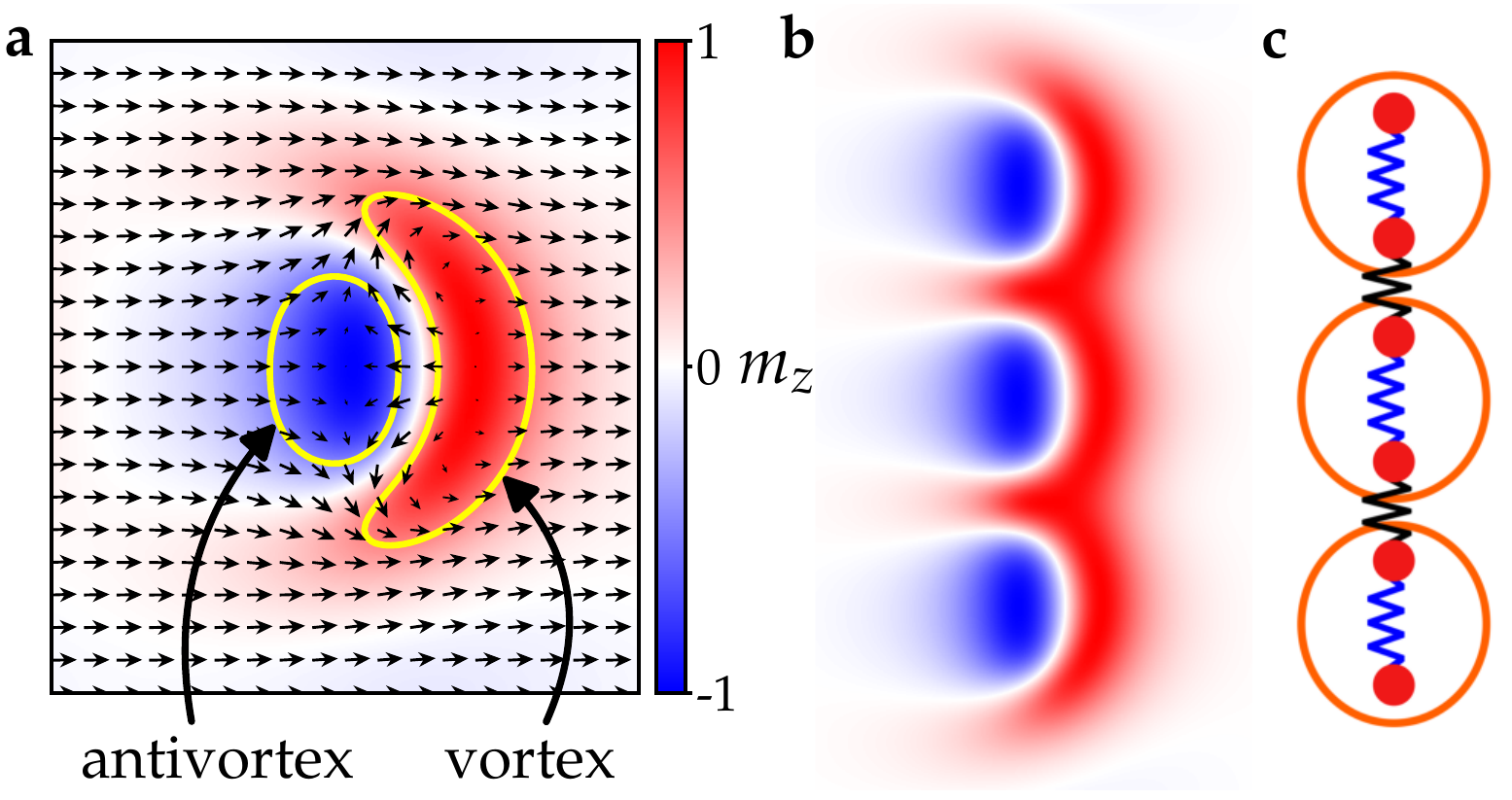}
	\caption{{\bf Asymmetric antibimerons (AABs) and their effective mechanical representation.} {\bf a} Magnetization configuration of a single AAB. {\bf b} AAB cluster containing $N=3$ antibimerons. In panels ({\bf a}) and ({\bf b}), colors indicate the out-of-plane magnetization component. {\bf c} Schematic of the corresponding one-dimensional spring-mass model describing AAB dynamics. Each AAB is represented as a meron dimer (outlined in orange) comprising two merons (red) coupled by an intra-dimer spring (blue), while neighbouring dimers are connected by inter-dimer springs (black).}
	\label{fig1}
\end{figure}

This knowledge gap is especially acute for in-plane magnetized meron-based textures~\cite{Kharkov2017, Gobel2019, Gao2019, Chen2025, Tretiakov2007}, which broaden the landscape of topological magnetism beyond conventional skyrmions. Among them, asymmetric antibimerons (AABs)~\cite{Vorobyev2024} constitute a distinctive class of composite textures formed by a bound antivortex--vortex pair, in which the vortex acquires a crescent shape (Fig.~\ref{fig1}a). Their broken rotational symmetry gives rise to a strongly anisotropic dynamics, while attractive interactions along one principal axis allow them to merge into clusters~\cite{Castro2025, Li20}, e.g. see Fig.~\ref{fig1}b for $N=3$ cluster. These features make AABs a particularly suitable system for investigating how localized eigenmodes of an individual asymmetric spin texture evolve into collective modes when several such objects are coupled together. AABs therefore provide an ideal platform for establishing a framework to classify modes in symmetry-broken topological systems.
 
Here, we combine micromagnetic simulations with spin-wave theory to identify the low-lying spin-wave modes of isolated AABs and their clusters in ultrathin ferromagnetic (FM) films under ac magnetic fields. We demonstrate that an isolated AAB exhibits a discrete set of localized eigenmodes which, upon cluster formation, split into multiplets whose number of components is determined by the cluster size. To elucidate this behavior, we introduce a topology-inspired coupled-oscillator model in which each AAB is mapped onto its constituent meron pair (Fig.~\ref{fig1}c). The resulting mode-resolved spectral fingerprints offer a practical route to identify complex bound-state configurations within the emerging class of asymmetric meron-based textures, which have recently been realized experimentally~\cite{Ohara2022, Yu2024}. Our framework enables unambiguous decoding of ferromagnetic resonance~\cite{Aliev2009} and Brillouin light scattering~\cite{Vogt2011} spectra, providing structural insight into complex magnetization configurations even when they remain inaccessible to direct imaging. Furthermore, the classification developed in this work is readily transferable to a broader class of asymmetric meron-based textures, providing a general foundation for their spectral characterization.

\section*{Results}

\subsection*{Micromagnetic model} We consider a thin FM film with periodic boundary conditions in its in-plane dimensions. AABs are stabilized by an interplay of the isotropic exchange, Dzyaloshinskii-Moriya interaction (DMI), uniaxial in-plane magnetic anisotropy, and constant external in-plane magnetic field.
Assuming slowly varying magnetization, the magnetic free energy of the system can be given as
\begin{equation}
F[\mathbf{m}]= d \int  d^2r \left[A(\mathbf{\nabla} \mathbf{m})^2 + \epsilon_{\mathrm{a}} + \	\epsilon_{\mathrm{DM}} - M_{\mathrm{s}}\mathbf{m}\cdot(\mathbf{B}+\mathbf{B}') \right],
\label{eq1}
\end{equation}
where $A > 0$ is the exchange constant, $\mathbf{m}$ is the normalized magnetization,  $\epsilon_{\mathrm{a}}=K_{x}[1-(\mathbf{m}\cdot  \mathbf{e}_{x})^2]$ is the easy-axis in-plane anisotropy energy density with anisotropy constant $K_{x}>0$, $\epsilon_{\mathrm{DM}}=D \left(\mathbf{e}_x \cdot \mathbf{m} \times \partial_y\mathbf{m} + \mathbf{e}_y \cdot \mathbf{m} \times \partial_x\mathbf{m}\right)$ is the DMI energy density with the DMI constant $D$, $\mathbf{B}$ is the static magnetic field, $\mathbf{B}'$ is the applied time-dependent magnetic field, $M_{\mathrm{s}}$ is the saturation magnetization, and $d$ is the film thickness.

Magnetization dynamics is simulated by the Landau–Lifshitz–Gilbert equation, 
\begin{equation}
\frac{d\vm}{dt} = \gamma_0 \mathbf{H}_{\rm eff} \times \vm + \alpha \vm \times \frac{d\vm}{dt},
\label{eq:LLG}
\end{equation}
where $\alpha$ is the dimensionless Gilbert damping constant, $\gamma_0$ is the gyromagnetic ratio and the effective magnetic field is $\mathbf{H}_{\rm eff} = - (\mu_{0} M_{\mathrm{s}})^{-1} \delta F[\mathbf{m}] / \delta \mathbf{m}$.

\subsection*{Spectral analysis} In order to identify the eigenmodes of AABs, we excite the system by a uniform low-amplitude magnetic field pulse with a cutoff frequency $f_{\mathrm{c}} = 50$ GHz. The temporal profile of the pulse is taken as a cardinal sine, $\mathrm{sinc}(2\pi f_{\mathrm{c}} t)=\sin(2\pi f_{\mathrm{c}} t)/(2\pi f_{\mathrm{c}} t)$. Here, we apply a magnetic field linearly polarized along the $x$-axis, parallel to the uniform background magnetization. In a perfectly symmetric texture, each eigenmode is typically excited only by a particular polarization. However, broken circular symmetry in an asymmetric texture such as an AAB hybridizes the modes. Consequently, each eigenmode in our study is a superposition of polarization-selective modes, enabling a single polarization to excite multiple eigenmodes simultaneously. 

We monitor the resultant AAB dynamics by evaluating the local magnetization fluctuations $ \delta {\mathbf{m}}({\mathbf{r}}, t)= {\mathbf{m}}({\mathbf{r}}, t) - {\mathbf{m}}({\mathbf{r}}, t=0)$, where ${\mathbf{m}}({\mathbf{r}}, t=0)$ is the initial (equilibrium) magnetization at each spatial point ${\mathbf{r}}$. Next, we perform a 2D Fourier transformation in real space $(x,y)$ to obtain the momentum-resolved fluctuations, $\delta {\mathbf{m}}_{k_x, k_y}(t) = \sum_{x} \sum_{y} \delta {\mathbf{m}}({\mathbf{r}}, t) \, e^{-i(k_x x + k_y y)}$. To quantify the overall strength of the induced response, we compute the frequency-domain power spectrum $S(f)$ as $ S(f) = | \int_{0}^{t_0} A(t) e^{-i2\pi f t} dt |^2$, where $t_{0}$ is the total duration of the time integration performed in the simulations. Here, we construct the momentum-integrated excitation amplitude as \(A(t) = \sum_{k_x}\sum_{k_y} | \delta {\mathbf{m}}_{k_x, k_y}(t)|^2\). A rigid translation of the texture produces no change in the net magnetization when comparing the states before and after the excitation, despite the spatial displacement of the texture. However, the local magnetization does change, and this displacement is encoded in the spatially nonuniform (finite-momentum) part of the response, which is captured by the momentum-space integration in $A(t)$. The amplitude $A(t)$ therefore provides a robust scalar measure of the total dynamical response of the magnetic system, which is excited by this pulse.

We distinguish localized AAB modes from the magnon continuum by comparing our spectra with the analytically derived magnon dispersion (see Supplementary Information for details). The continuum is defined by the magnon dispersion of the uniform FM background. The two field orientations are treated independently: the in-plane and out-of-plane geometries correspond to different equilibrium configurations and hence different magnon continua. We therefore compare each spectrum with the corresponding analytical dispersion. In the presence of an in-plane field, the magnon dispersion is given by
\begin{equation}
\omega_{\mathbf{k}} = \frac{\gamma_0}{\mu_0 M_{\mathrm{s}}} \Biggl[
\frac{4A}{a^2} \bigl( 2 - \cos k_x - \cos k_y \bigr) 
+ \frac{2D}{a}\,\sin k_y 
+ 2K + M_{\mathrm{s}} B_x \Biggr],
\label{eq6}
\end{equation}
where $\mu_0$ is the vacuum permeability, and $a$ is the lattice constant. For an out-of-plane field, the magnon dispersion becomes
\begin{equation}
\omega_{\mathbf{k}}
= \frac{\gamma_0}{\mu_0 M_{\mathrm{s}}}\sqrt{\epsilon_{\mathbf{k}}^2-\Delta_{\mathbf{k}}^2},
\label{eq:omega_outplane_main}
\end{equation}
with
\begin{align*}
\epsilon_{\mathbf{k}}
&=
\frac{4A}{a^2}\bigl(2-\cos k_x-\cos k_y\bigr)
+ \frac{2D}{a}\cos\theta\,\sin k_y
\\[4pt]
&\quad
+ 2K\cos^2\theta
- K\sin^2\theta
- M_{\mathrm{s}} B_z \sin\theta
\end{align*}
and $\Delta_{\mathbf{k}} = - K\sin^2\theta$. Here, $\theta$ is the canting angle of the equilibrium magnetization away from the film's plane induced by the applied field $B_z$. In what follows, localized AAB modes are identified as discrete spectral peaks lying below the magnon continuum defined by Eq.~\eqref{eq6} or Eq.~\eqref{eq:omega_outplane_main}, respectively.
 
\subsection*{Isolated AABs} 
\begin{figure}[!htb]
	\centering
	\includegraphics[width=1\linewidth]{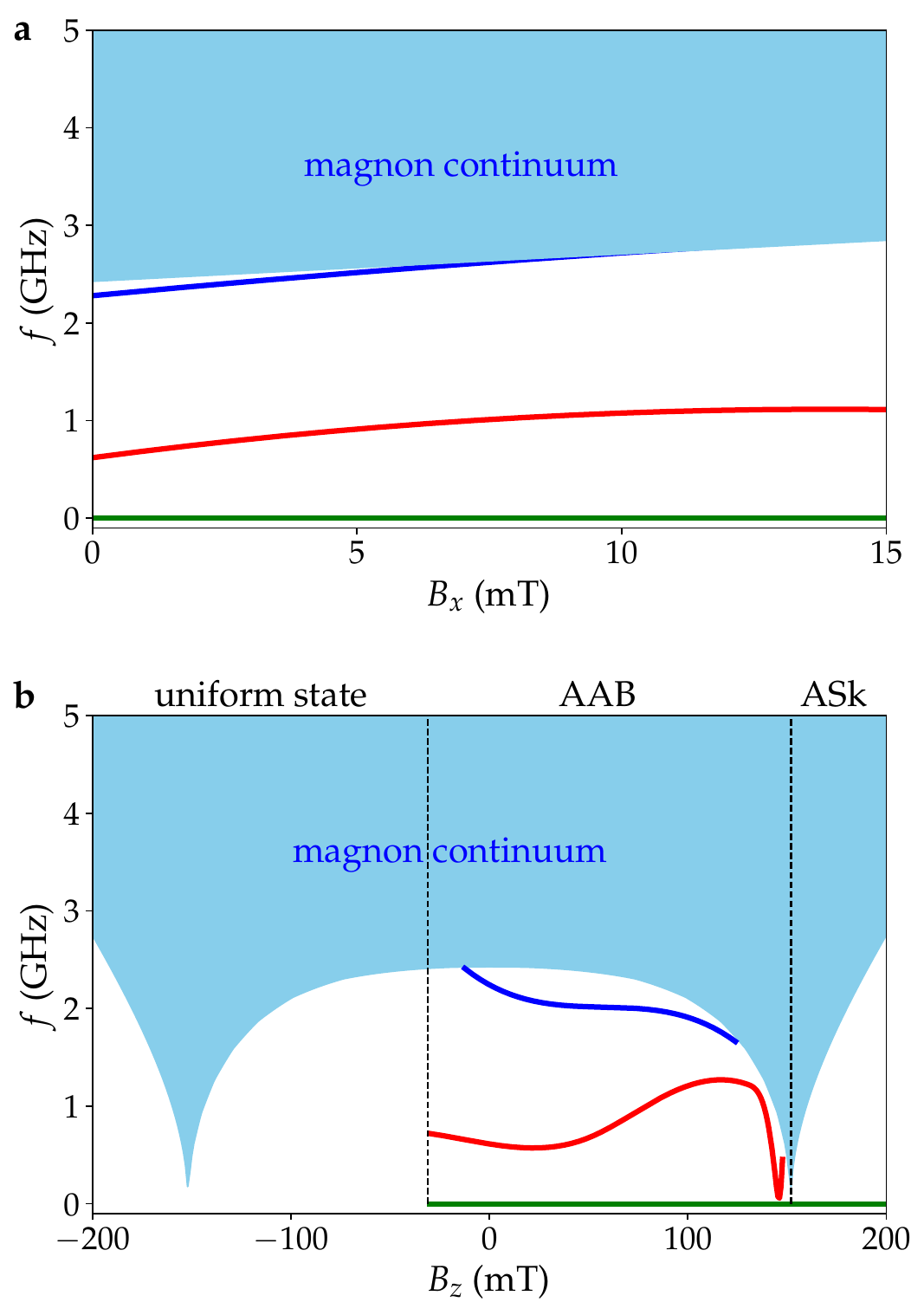}
	\caption{{\bf Localized eigenmode spectra of an asymmetric antibimeron with topological charge $Q=+1$ as a function of external magnetic field.} Panel ({\bf a}) shows the in-plane field $B_x$ dependence, and ({\bf b}) shows the out-of-plane field $B_z$ dependence. The shaded blue region corresponds to the magnon continuum.}
	\label{fig2}
\end{figure}

Using the spectral analysis described above, we calculate the power spectra of an isolated AAB as a function of static magnetic field applied either in-plane (\(B_x\)) or out-of-plane (\(B_z\)), as summarized in Fig.~\ref{fig2}. Unless stated otherwise, we use the following material parameters: \(M_{\mathrm{s}}=111 \,\mathrm{kA \, m^{-1}}\), \(A=1.25 \,\mathrm{pJ \, m^{-1}}\), \(D=0.135 \,\mathrm{mJ \, m^{-2}}\), and \(K=8.5 \,\mathrm{kJ \, m^{-3}}\). Varying \(B_x\) modifies the AAB, which can be stabilized both at \(B_x=0\) and at finite \(B_x\), allowing us to track how its localized eigenmodes evolve as field \(B_x\) deforms the texture. An out-of-plane field \(B_z\), in turn, continuously deforms the spin configuration from an in-plane AAB into an antiskyrmion. Therefore, sweeping \(B_z\) enables us to map the mode evolution across this AAB-to-antiskyrmion crossover within a single framework. A static in-plane field \(B_y\) yields no qualitatively new behavior and thus its analysis is omitted for brevity. Since external static fields can be readily tuned in experiments, these parameter sweeps provide a direct way to reproduce our results.

Figure~\ref{fig2}a shows the power spectrum of the AAB state with topological charge $Q = \frac{1}{4\pi} \int \mathbf{m}\cdot(\partial_x \mathbf{m} \times \partial_y \mathbf{m}) dxdy =+1$ as a function of  $B_x$. Three localized modes appear below the bottom of the magnon continuum. The lowest is the zero mode (denoted as $Z$ mode), associated with the translation of the undistorted AAB and guaranteed by the translational symmetry of Eq.~\eqref{eq1}. It is indicated by the green curve in Fig.~\ref{fig2}a. In practice, perturbations that break translational symmetry (e.g., impurities) can weakly hybridize the translational mode with the elongation mode and lift it to a finite excitation frequency.

In the special case of a circularly symmetric texture, such as a skyrmion, the eigenmodes can be classified by an integer azimuthal quantum number $\mu$~\cite{Kravchuk2019, Mochizuki2012, Garst2014} (e.g., $\mu=0$ for the breathing mode and $\mu=\pm 1$ for the gyrotropic modes). In contrast, the intrinsic asymmetry of the AAB breaks rotational symmetry, so modes hybridize and can no longer be assigned a well-defined $\mu$. In particular, instead of $\mu=0$ breathing mode with radial symmetry, the AAB exhibits an elongation mode in which the texture periodically stretches and compresses predominantly along a fixed axis (the $y$-axis). Thus, while a symmetric skyrmion supports a purely radial “breathing” excitation, the AAB response is dominated by elongation, but includes a small gyrotropic component. With this in mind, we identify the second spectral peak as an elongation ($E$) mode.

\begin{figure}[!htb]
	\centering
	\includegraphics[width=1\linewidth]{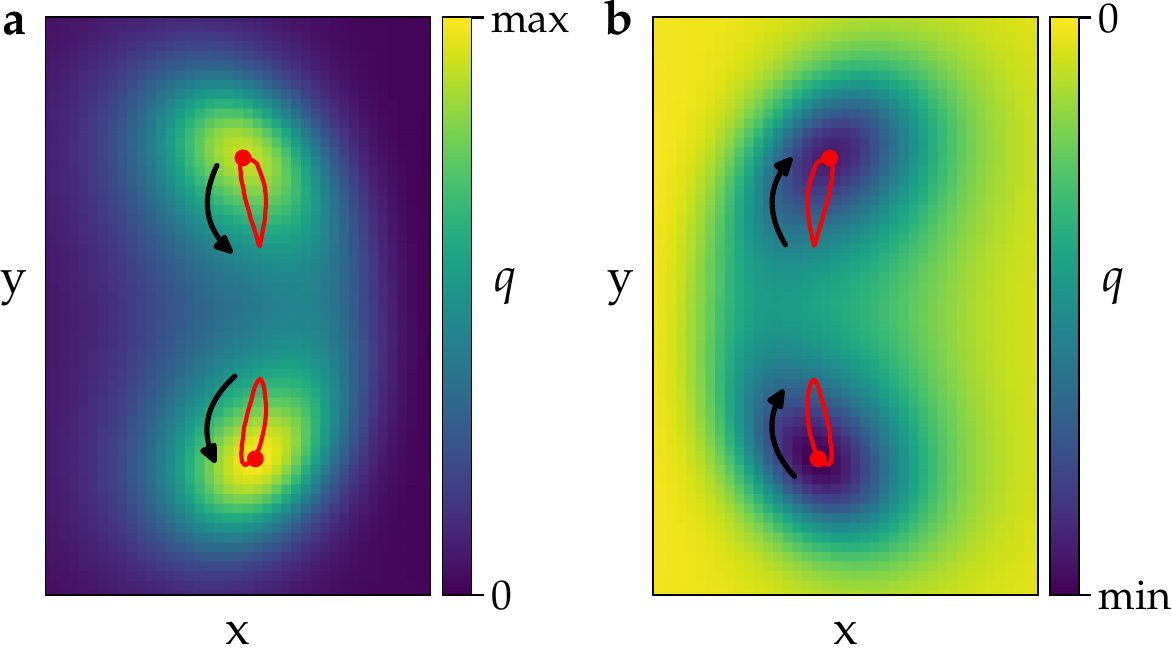}
	\caption{{\bf Gyration of the topological charge density peaks in asymmetric antibimerons (AABs).} Topological charge density distribution for AAB with ({\bf a}) $Q$=+1 ({\bf b}) $Q$=-1. Red dots and loops represent the minima ({\bf a}) and maxima ({\bf b}) of the topological charge, along with their corresponding trajectories over one period of AAB oscillation. Reversing the topological charge ($Q\to -Q$) inverts the density, $q(\mathbf r)\to -q(\mathbf r)$, such that $q_{\max}^{(a)} = \lvert q_{\min}^{(b)}\rvert$. Black arrows indicate the direction of rotation associated with the gyrotropic mode.}
	\label{fig3}
\end{figure}

The $E$ mode appears as a resonance peak in the range \(f \simeq 0.6\,\text{--}\,1.1\,\mathrm{GHz}\) throughout the entire \(B_x\) sweep, see the red curve in Fig.~\ref{fig2}a and Supplementary Movie 1. 
In this mode, the gyrotropic component exhibits an interesting feature. Because the AAB lacks rotational symmetry, the topological charge density, \( q = [\mathbf{m}\cdot(\partial_x \mathbf{m} \times \partial_y \mathbf{m})]/4\pi \), is distributed asymmetrically and forms two pronounced peaks, as shown in Fig.~\ref{fig3}. The gyrotropic component then manifests as a rotation of these peaks around the out-of-plane axis, with the sense of rotation set by the topological charge: counterclockwise for $Q=+1$ and clockwise for $Q=-1$ (Fig.~\ref{fig3}).

The third resonance peak appears just below the bottom of the magnon continuum, at $f \simeq 2.2$ GHz for $B_x=0$, see blue curve in Fig.~\ref{fig2}a and Supplementary Movie 2.
Since it lies close to the continuum edge, this excitation strongly hybridizes with delocalized magnon modes. Because this excitation is dominated by them, we denote this resonance as $M$ mode. This interpretation is supported by the analytical model introduced below, which reproduces the discrete $Z$ and $E$ eigenmodes of an isolated AAB but yields no additional discrete mode corresponding to this resonance. The frequency of this mode increases more steeply with $B_x$ than the lower edge of the continuum, causing them to merge at \(B_x \simeq 7\)\,mT. We note that the full power spectrum of an AAB with $Q=-1$, including the $Z$, $E$, and $M$ modes, coincides with that for $Q=+1$ shown in Fig.~\ref{fig2}a. The reason is that AABs with opposite topological charges are related by a mirror symmetry of the energy functional. As a result, the spin textures with $Q=\pm1$ are energetically degenerate~\cite{Vorobyev2024} and therefore have the same eigenmode spectrum.

Panel~(b) of Fig.~\ref{fig2} shows the power spectrum for the AAB state with $Q=+1$ as a function of the $B_z$ field. The $Z$, $E$, and $M$ modes are shown in green, red, and blue, respectively, as in panel~(a). As $B_z$ increases, the AAB continuously transforms into a symmetric ASk, accompanied by a smooth evolution of the mode frequencies. The dashed line at $B_z \simeq 152\,\mathrm{mT}$ marks the point beyond which the in-plane component of the background magnetization vanishes, and for $B_z$ above this value the spectrum corresponds to that of an ASk. When $B_z$ is swept in the negative direction, once the field exceeds the threshold $B_z \simeq -31\,\mathrm{mT}$, the metastable AAB state decays into energetically more favorable uniformly magnetized state. The system in this state behaves like a single-domain macrospin, giving rise to a single FM resonance mode. The corresponding eigenmode spectrum for an AAB with $Q=-1$ is the mirror image under $B_z\!\to\!-B_z$ of the $Q=+1$ dependence shown in Fig.~\ref{fig2}b, reflecting the symmetry relating AABs with opposite topological charges. An out-of-plane field lifts the $Q=\pm1$ degeneracy and increases the energy splitting between the $Q=+1$ and $Q=-1$ states~\cite{Vorobyev2024}. Overall, these topological transitions demonstrate how systematic variation of $B_z$ tunes the system between AAB, antiskyrmion, and the uniform ground state, thereby modifying the observed spin-wave modes. 

\subsection*{AAB clusters} 
\begin{figure}[!htb]
	\centering
	\includegraphics[width=1\linewidth]{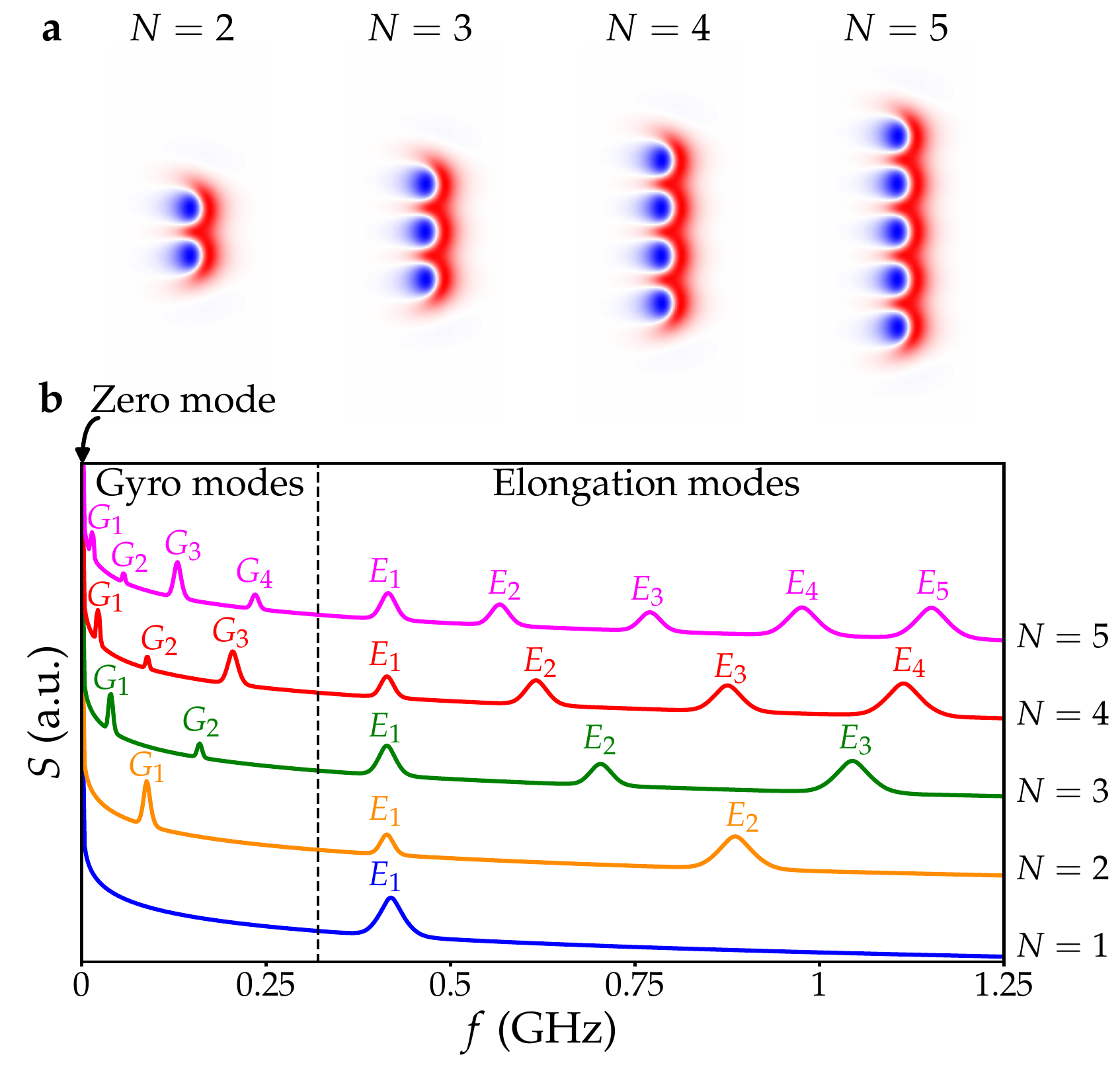}
	\caption{{\bf Asymmetric antibimeron (AAB) clusters and their power spectra.} {\bf a} Magnetization configurations of AAB clusters with $N$ ranging from 2 to 5. The colors indicate the out-of-plane magnetization component. {\bf b} Power spectra $S (f)$ of AAB clusters. The bottom curve represents an isolated AAB for comparison, while the remaining colored curves correspond to the clusters containing 2 to 5 AABs.}
	\label{fig4}
\end{figure}

Clusters of skyrmions and bimerons have been predicted and observed in chiral FMs~\cite{Loudon2018}, AFMs~\cite{Li20}, frustrated materials~\cite{Naya2022}, and liquid crystals~\cite{Sohn2019}. Unlike conventional skyrmions, AABs with the same topological charge exhibit attractive interaction perpendicular to the easy-axis anisotropy direction, enabling them to merge into clusters with a wide range of $Q$ values. This clustering gives rise to collective spin excitations that are qualitatively different from those of isolated skyrmions and bimerons. Here, we study power spectra of AAB clusters (Fig.~\ref{fig4}) in the same FM film as previously used for the isolated AAB case. For the cluster simulations, we employ a reduced exchange stiffness \(A=0.69 \, \mathrm{pJ \, m^{-1}}\) and fix the in-plane field at \(B_x=40\,\mathrm{mT}\). All other parameters are kept the same as in the isolated-AAB simulations.

The simulated power spectra of AAB clusters are shown in Fig.~\ref{fig4}b. For our analysis, we focus exclusively on the zero $Z$ and elongation $E$ modes, as they are well separated from the continuum. The $M$ mode, located just below the magnon continuum, partially merges into it due to mode splitting and is therefore not examined in detail. Moreover, the modes near the continuum are dominated by magnons that are not localized on the AAB clusters and are thus beyond the scope of this work. As the number $N$ of AABs in the cluster increases, so does the total number of modes. Each isolated-AAB $Z$ or $E$ mode splits into multiple modes, with the zero mode remaining and additional gyrotropic ($G_i$) and elongation ($E_i$) modes emerging. For clarity, we split the spectra into two frequency ranges, one for the gyrotropic and the other one for the elongation modes. The $Z$ mode and elongation $E_{1}$ mode remain nearly constant for all $N$, whereas higher-order modes, such as $G_{1}$, $E_{2}$, and others, progressively shift to lower frequencies with increasing cluster size (Fig.~\ref{fig4}b). Thus, the constant frequencies of $Z$ and $E_{1}$ modes reflect their uncoupled character, whereas the downward shift in the other modes indicates the increasing role of inter-AAB coupling in the cluster dynamics.

We also note that, while an isolated AAB exhibits hybridization among its modes, the relevant eigenfrequencies are well separated, so each resonance retains a dominant character and can be classified unambiguously. In clusters, by contrast, each isolated-AAB mode splits into a closely spaced multiplet. The resulting small separations between neighboring modes enhance their hybridization, yielding eigenmodes with appreciable gyrotropic and elongation contributions. Nevertheless, $G_i$ and $E_i$ labels remain meaningful because each resonance is still dominated by either gyrotropic or elongation character, as demonstrated by the analytical model introduced below.

We analyze two representative cases in detail, an even cluster ($N=2$) and an odd cluster ($N=3$), and characterize each mode by its time-dependent magnetization dynamics. For $N=2$, each isolated-AAB mode forms a doublet: $Z$ mode splits into $Z$ and $G_1$, whereas the elongation mode $E$ splits into $E_1$ and $E_2$, as shown by the orange curve in Fig.~\ref{fig4}b. The $Z$ mode is the translational Goldstone mode, i.e., a rigid in-phase oscillation of all AABs in the cluster. The $G_{1}$ mode is dominated by a gyrotropic excitation in which the two AABs precess out-of-phase (Supplementary Movie 3). 
In this mode, each AAB remains nearly rigid, with only weak internal deformation, but the out-of-phase motion modulates the AABs mutual overlap region. The elongation branch shows an analogous in-phase/out-of-phase structure. The lower-frequency mode $E_1$ corresponds to predominantly out-of-phase elongation of the two AABs (Supplementary Movie 4). 
The textures elongate with negligible displacement of their centers from equilibrium, so that the AABs mutual overlap region remains nearly unchanged. By contrast, in $E_2$ the elongations are nearly in phase (Supplementary Movie 5), 
so that the cluster expands and contracts coherently. Accordingly, in $E_1$ the total cluster length is weakly modulated, whereas in $E_2$ cluster expansion and contraction are pronounced.

For $N=3$ cluster, the isolated-AAB modes split into triplets: the zero mode yields $Z$, $G_1$, and $G_2$, whereas the elongation mode yields $E_1$, $E_2$, and $E_3$, as shown by the green curve in Fig.~\ref{fig4}b. The $Z$ mode remains at zero frequency and corresponds to a rigid in-phase translation of all three spin textures. In the $G_{1}$ mode, the central AAB remains comparatively stationary, while the edge AABs provide the dominant gyrotropic dynamics (Supplementary Movie 6). 
The two edge spin textures undergo out-of-phase precession, closely resembling the $G_{1}$ mode of the $N=2$ cluster. In the $G_2$ mode, the edge AABs precess nearly in-phase with each other but out-of-phase with the central AAB, and the gyrotropic component is reduced compared with $G_1$ (Supplementary Movie 7). 
The elongation branch again forms a set of modes distinguished by their relative phase relations. In the $E_1$ mode, the two edge AABs elongate nearly in-phase, whereas the central AAB elongates out-of-phase with them (Supplementary Movie 8). 
In the $E_2$ mode, the two edge AABs elongate approximately out-of-phase, while the central texture exhibits a weaker elongation response (Supplementary Movie 9). 
Finally, in the $E_3$ mode, all three AABs elongate nearly in-phase, producing the strongest coherent modulation of the cluster length within the elongation branch (Supplementary Movie 10).

\subsection*{Coupled-oscillator model for AAB clusters} 
\begin{figure}[!htb]
	\centering
	\includegraphics[width=1\linewidth]{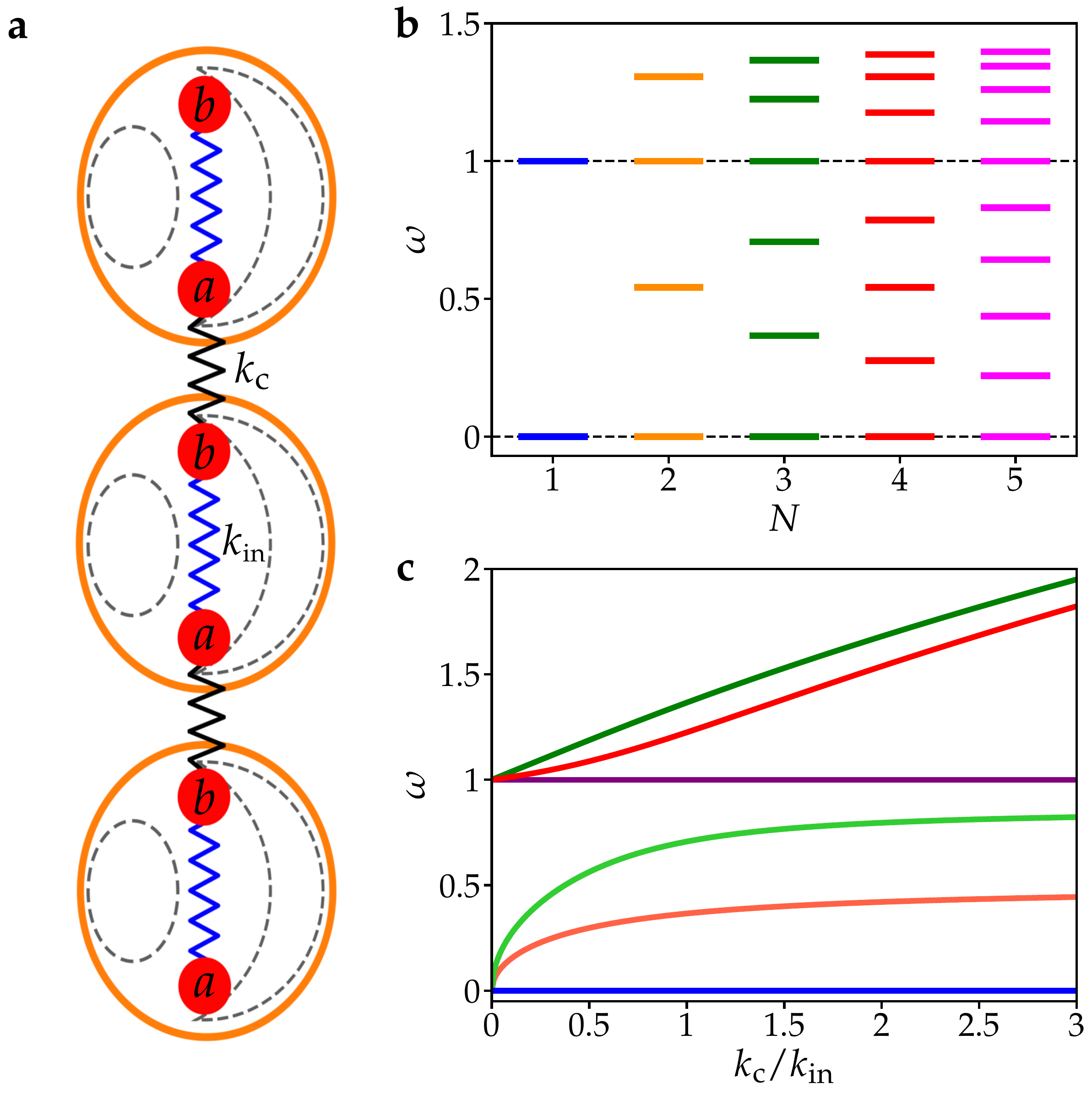}
	\caption{{\bf Mechanical representation of asymmetric antibimeron (AAB) clusters and corresponding normal-mode spectra.} {\bf a} Spring-mass schematic of the meron-dimer model for an AAB cluster with $N=3$. Each AAB is modeled as two particles (corresponding to merons) of mass $m$, labeled $a$ and $b$, connected by an intra-dimer spring with stiffness $k_{\mathrm{in}}$. Adjacent dimers are coupled by an inter-dimer spring with stiffness $k_{\mathrm{c}}$. Grey dashed contours schematically indicate the underlying spin textures. {\bf b} Normal-mode spectra of the coupled-oscillator model for clusters containing up to $N=5$ dimers, shown for $k_{\mathrm{in}}=k_{\mathrm{c}}$. {\bf c} Dependence of the eigenmode frequencies on the coupling ratio $k_{\mathrm{c}}/k_{\mathrm{in}}$ for a three-dimer ($N=3$) chain. In panels ({\bf b}) and ({\bf c}), the frequency $\omega$ is expressed in units of $\omega_1 = \sqrt{2 k_{\mathrm{in}} / m}$.}
	\label{fig5}
\end{figure}

To elucidate the origin of the observed mode splitting and the role of inter-AAB coupling, we develop a coupled-oscillator model that captures the essential features of the low-energy collective excitations in AAB clusters. In previous work~\cite{Vorobyev2024}, we approximated the anisotropic current-driven dynamics of an isolated AAB by treating its two constituent merons separately; however, this approach is insufficient for describing AAB excitations. The centers of these merons correspond to the two peaks of the topological charge density within the AAB (see, e.g. Fig.~\ref{fig3}). Motivated by this meron-based picture, we map AAB clusters onto a spring-mass mechanical system, in which the constituent merons are treated as particles of effective mass $m$ coupled by springs (Fig.~\ref{fig5}a). We represent each AAB as a topological dimer composed of two merons, labeled $a$ and $b$, and model an AAB cluster as a one-dimensional chain of $N$ such dimers. The intra-dimer interaction (within a single AAB) is modeled by a spring of stiffness $k_{\mathrm{in}}$. The inter-dimer interaction (between adjacent AABs) is modeled by a spring with stiffness $k_{\mathrm{c}}$, which connects particle $b$ of the $i$-th dimer to particle $a$ of the $(i+1)$-th dimer. 

In the small-oscillation regime, we describe the dynamics of an AAB cluster in terms of small displacements of merons from their equilibrium positions, $\xi_{i}^a$ and $\xi_{i}^b$, where $i=1,\ldots,N$. Within this mechanical analogy, the AAB chain is described by the Lagrangian:

\begin{equation}
\label{fig:L}
\mathcal{L} = \frac{m}{2}\sum_{i=1}^N \left[ ( \dot{\xi}_{i}^a)^2 + ( \dot{\xi}_{i}^b)^2 \right]
-\frac{k_{\mathrm{in}}}{2}\sum_{i=1}^N{(\xi_{i}^a - \xi_{i}^b)}^2
-\frac{k_{\mathrm{c}}}{2}\sum_{i=1}^{N-1}{(\xi_{i+1}^a - \xi_{i}^b)}^2 \! .
\end{equation}
By solving the equations of motion derived from the Lagrangian given by Eq.~(\ref{fig:L}) for a chain of $N$ dimers, we obtain the normal frequencies shown in Fig.~\ref{fig5}b. Two normal modes occur at fixed frequencies $\omega_0=0$ and $\omega_1=\sqrt{2k_{\mathrm{in}}/m}$, indicated by the black dashed lines in Fig.~\ref{fig5}b. Note that these frequencies are independent of chain length. The zero mode at $\omega_0$ corresponds to a rigid translation of the entire chain. The mode with $\omega_1$ is set solely by the intra-dimer spring and therefore corresponds to an internal dimer oscillation. In the expression for the frequency of this mode, the factor of 2 arises because the two particles of equal-mass oscillate out-of-phase, so their relative motion is governed by an effective stiffness $2k_{\mathrm{in}}$. For $N=1$, the eigenmodes with $\omega_0$ and $\omega_1$ correspond to $Z$ and $E$ modes of an isolated AAB, reproducing rigid translation and elongation dynamics, respectively.

For $N>1$, all other eigenmodes split into lower- and higher-frequency branches separated by $\omega_1$. Figure~\ref{fig5}c shows the evolution of these branches with the coupling ratio $k_{\mathrm{c}}/k_{\mathrm{in}}$ for $N=3$ chain depicted in Fig.~\ref{fig5}a. The low-frequency branch ($\omega<\omega_1$) is governed primarily by inter-dimer coupling. As $k_{\mathrm{c}}$ increases, the frequencies in this branch saturate at finite values. This saturation occurs because strong inter-dimer coupling effectively forces adjacent dimers to move together. Consequently, the dimers oscillate as nearly rigid units, so the mode frequencies in this branch are governed primarily by the weaker intra-dimer springs with $k_{\mathrm{in}}$. The higher-frequency branch ($\omega>\omega_1$), by contrast, corresponds to internal deformations of the dimers, and the frequencies in this branch increase with $k_{\mathrm{c}}$, since stronger inter-dimer coupling increasingly penalizes such distortions. Ultimately, this spectral separation firmly substantiates our classification of AAB modes into gyrotropic $G$ and elongation $E$ modes.

This topological coupled-oscillator model reproduces the micromagnetic results remarkably well, both in the number of eigenmodes and in the hierarchy of eigenfrequencies across different AAB cluster sizes, see Figs.~\ref{fig4}b and \ref{fig5}b. In addition to the two modes with $\omega_0$ and $\omega_1$, the model predicts an extra pair of eigenmodes $\omega_{2,3} = [k_{\mathrm{c}} + k_{\mathrm{in}} \pm (k_{\mathrm{c}}^2 + k_{\mathrm{in}}^2)^{1/2}]^{1/2}/\sqrt m$ whose frequencies are independent of cluster size, for all clusters with even $N$. In the micromagnetic simulations, these modes are identified as $G_1$ and $E_2$ for $N=2$ cluster, and as $G_2$ and $E_3$ for $N=4$ cluster, see Fig.~\ref{fig4}b.
We animate the corresponding normal modes (see Supplementary Movie 11) 
of the coupled-oscillator model for chains up to $N=3$ in the right panels of Supplementary Movies 1 and 3--10, 
enabling direct visual comparison with the corresponding micromagnetic simulations. Strikingly, despite its one-dimensional character explicitly implying the absence of meron gyration~\cite{Suppl12}, the coupled-oscillator model reproduces the dynamics observed in the micromagnetic simulations qualitatively across all modes represented in the Supplementary Movies. In particular, it captures the same phase relations between neighbouring AABs and the same characteristic deformation patterns. Overall, the excellent agreement between the micromagnetic simulations and the model demonstrates that the collective spectra are primarily governed by the interplay between intra- and inter-dimer interactions, which is robustly captured by our topology motivated coupled-oscillator model.

\section*{Discussion}

We also emphasize that all results presented in this work apply equally and without modification to asymmetric bimerons and their clusters~\cite{Ohara2022, Yu2024}, which are stabilized by a different type of DMI, $D\,(\mathbf{e}_y \cdot \mathbf{m} \times \partial_x\mathbf{m} - \mathbf{e}_x \cdot \mathbf{m} \times \partial_y\mathbf{m})$, than that responsible for the stabilization of AABs. Although the specific DMI symmetry determines the equilibrium spin configuration, it does not change the underlying topological structure of the texture, which in both cases consists of a bound pair of merons with a finite separation. As a result, the low-energy dynamics are governed by the same set of collective coordinates describing the relative displacement and deformation of the constituent merons. This correspondence is not merely qualitative: we find that the excitation spectra of asymmetric bimerons and AABs completely coincide, including the full mode hierarchy and its evolution with cluster size (see Supplementary Fig. 1).

More generally, the physical framework developed here extends beyond in-plane magnetized textures to a broader class of asymmetric meron-based spin textures. In particular, while the DMI considered in Ref.~\cite{Gobel2019} is known to stabilize symmetric bimerons, the same systems can host asymmetric textures once an out-of-plane anisotropy is present and the net magnetization acquires an out-of-plane component and is no longer perpendicular to all DMI vectors. Under these conditions, asymmetric (anti-)skyrmions can be stabilized, which share the same meron-pair topology that governs the dynamics discussed in this work. Consequently, their low-energy excitations can be described within an equivalent normal-mode framework. This demonstrates that the dynamical concepts introduced here are not limited to a specific texture or stabilizing interaction, but instead apply broadly to asymmetric meron-based spin textures.

In summary, we have theoretically investigated the spin-wave modes of AABs and their clusters in ultrathin FM films. For isolated AABs, we identified a discrete spectrum of localized modes, including a zero-frequency translational mode associated with rigid motion of the spin texture, an elongation mode corresponding to internal shape deformation, and a higher-energy resonance strongly hybridized with the magnon continuum. Furthermore, we analyzed the effects of both in-plane and out-of-plane magnetic fields on the localized excitation spectra. In particular, we showed that an applied out-of-plane field drives a continuous evolution of the eigenmodes as the AAB smoothly transforms into an antiskyrmion. This evolution across the topological transition highlights the fundamental interplay between symmetry, topology, and excitation spectra in spin textures.

For clusters comprising $N$ AABs, we show that discrete localized resonances of a single AAB split into $N$-fold multiplets. This splitting leaves the translational Goldstone mode $Z$ and the lowest elongation mode $E_1$ nearly insensitive to cluster size, but higher-order modes soften systematically with increasing $N$, see Fig.~\ref{fig4}b. To understand this mode evolution, we developed a topology-constrained coupled-oscillator model based on interacting meron dimers, which reproduces the observed mode splitting and frequency ordering. Within this framework, the collective excitations are described in terms of well-defined normal modes, with their total number equal to the number of localized degrees of freedom (i.e., the number of merons). Moreover, this model justifies the mode classification we developed based on micromagnetic simulations and supports the separation of the cluster spectra into gyrotropic- and elongation-dominated branches. Overall, our results demonstrate that the collective low-energy dynamics of asymmetric meron-based textures is governed by the interplay of topology and inter-texture coupling. These findings establish AAB clusters as reconfigurable multimode spin-wave nano-oscillators~\cite{Garcia-Sanchez_2016, Shen2019, Shen2026}, enabling mode-selective excitation that offers a route to nanoscale magnonic signal-processing circuits~\cite{Flebus2024}, including logic and neuromorphic computing networks~\cite{Grollier2020, Marrows2024}.

\section*{Acknowledgments} The authors are grateful to Kei Yamamoto for valuable discussions. O.A.T. acknowledges support from the Australian Research Council (Grant No. DP240101062) and the NCMAS grant.

\section*{Author contributions} These authors contributed equally: Pavel A. Vorobyev and Daichi Kurebayashi.

\section*{Data availability} The data that support the findings of this study are available from the corresponding authors upon reasonable request. 

\section*{Competing interests} The authors declare no competing interests.

\bibliography{spin_wave_antibimeron}

\begin{thebibliography}{63}
\bibitem [{Sup()}]{Suppl12}%
  \BibitemOpen
  \bibinfo {note} {A gyrotropic term is not included explicitly, as its
  incorporation would require at least a two-dimensional collective-coordinate
  description of the meron dynamics. Such an extended description would be
  consistent with the gyration of the merons observed in the micromagnetic
  simulations, but lies beyond the scope of this work.}\BibitemShut {Stop}%
\end{thebibliography}

\onecolumngrid

\section*{Supplementary Information}

\setcounter{section}{0}
\setcounter{equation}{0}
\setcounter{figure}{0}
\setcounter{table}{0}

\renewcommand{\thesection}{S\arabic{section}}
\renewcommand{\theequation}{S\arabic{equation}}
\renewcommand{\thetable}{S\arabic{table}}

\makeatletter
\def\fnum@figure{Supplementary Fig.~\thefigure}
\makeatother

\section{Magnon dispersion}
\label{md}

\subsection{In-plane field}
\label{app:md_inplane}

We begin by considering the micromagnetic Hamiltonian given by Eq.~(1) in the main text to derive the magnon dispersion. This dispersion relation characterizes the magnon continuum excitations and enables us to distinguish them from the excitations corresponding to low-lying collective AAB modes. To recast this Hamiltonian in an atomistic framework, we introduce the following ladder operators:
\[
S_{\mathbf{r}}^+ = -S_{\mathbf{r}}^z + i S_{\mathbf{r}}^y, \quad S_{\mathbf{r}}^- = -S_{\mathbf{r}}^z - i S_{\mathbf{r}}^y. 
\]
Although these definitions appear unconventional, they are appropriate in our rotated frame where the magnetization is polarized along the $x$-axis. The spin components are taken to be dimensionless and obey the angular momentum commutation relations
\(
[S_i^\alpha,S_j^\beta]
= i\,\epsilon_{\alpha\beta\gamma} S_i^\gamma \delta_{ij}.
\)

With these ladder operators defined, we now turn to the atomistic formulation of the Hamiltonian. In this framework, the corresponding Hamiltonian reads:
\begin{equation}
\begin{aligned}[b]
H &= \frac{4AN^2}{a^2} 
+ \sum_{\mathbf{r}} \Bigg[
-\frac{2A}{a^2} \sum_{j \in \{x, y\}} 
\mathbf{S}_{\mathbf{r}} \cdot \mathbf{S}_{\mathbf{r} + \hat{e}_j} 
+ \frac{D}{a} \Big( 
S^z_{\mathbf{r}} S^x_{\mathbf{r} + \hat{e}_x} 
- S^x_{\mathbf{r}} S^z_{\mathbf{r} + \hat{e}_x} 
+ S^y_{\mathbf{r}} S^z_{\mathbf{r} + \hat{e}_y} 
- S^z_{\mathbf{r}} S^y_{\mathbf{r} + \hat{e}_y} 
\Big)
- K \left( S^x_{\mathbf{r}} \right)^2 
- M_{\mathrm{s}} B_x S^x_{\mathbf{r}} 
\Bigg] \\
&
= \frac{4AN^2}{a^2} 
+ \sum_{\mathbf{r}} \Bigg\{
-\frac{A}{a^2} \sum_{j \in \{x, y\}}
\Big( 
S^+_{\mathbf{r}} S^-_{\mathbf{r} + \hat{e}_j} 
+ S^-_{\mathbf{r}} S^+_{\mathbf{r} + \hat{e}_j} 
+ 2 S^x_{\mathbf{r}} S^x_{\mathbf{r} + \hat{e}_j} 
\Big)
+ \frac{D}{2a} \Bigg[
S^x_{\mathbf{r}} \left( S^+_{\mathbf{r} + \hat{e}_x} + S^-_{\mathbf{r} + \hat{e}_x} \right) 
- \left( S^+_{\mathbf{r}} + S^-_{\mathbf{r}} \right) S^x_{\mathbf{r} + \hat{e}_x} \\
&
+ i \left( S^+_{\mathbf{r}} S^-_{\mathbf{r} + \hat{e}_y} 
- S^-_{\mathbf{r}} S^+_{\mathbf{r} + \hat{e}_y} \right)
\Bigg] 
- K \left( S^x_{\mathbf{r}} \right)^2 
- M_{\mathrm{s}} B_x S^x_{\mathbf{r}} \Bigg\},
\label{eq2}
\end{aligned}
\end{equation}
where \(a\) is the lattice constant. To obtain the magnon dispersion, we expand around the fully polarized state (in the \(\hat{x}\)-direction) using the linearized Holstein--Primakoff transformation:
\[
S_{\mathbf{r}}^+ = \sqrt{2S} a_{\mathbf{r}}, \quad S_{\mathbf{r}}^- = \sqrt{2S} a_{\mathbf{r}}^\dagger, \quad S_{\mathbf{r}}^x = S - a_{\mathbf{r}}^\dagger a_{\mathbf{r}},
\]
where operator \(a_{\mathbf{r}}\) (\(a_{\mathbf{r}}^\dagger\)) annihilates (creates) a magnon localized at site \({\mathbf{r}}\). Retaining only bilinear terms in \(a_{\mathbf{r}}\) and \(a_{\mathbf{r}}^\dagger\) (i.e., the linear spin-wave approximation) maps Eq.~(\ref{eq2}) onto the spin-wave Hamiltonian
\begin{equation}
H_{\mathrm{SW}} = \sum_{\mathbf{r}} \Bigg[ 
-\frac{2SA}{a^2} \sum_{j \in \{x, y\}} \Big( a_{{\mathbf{r}}+\hat{e}_j}^\dagger a_{\mathbf{r}} + a_{\mathbf{r}}^\dagger a_{{\mathbf{r}}+\hat{e}_j} - 2a_{\mathbf{r}}^\dagger a_{\mathbf{r}} \Big) + \frac{iSD}{a} \Big( a_{{\mathbf{r}}+\hat{e}_y}^\dagger a_{\mathbf{r}} - a_{\mathbf{r}}^\dagger a_{{\mathbf{r}}+\hat{e}_y} \Big) + 2SK a_{\mathbf{r}}^\dagger a_{\mathbf{r}} + M_{\mathrm{s}} B_x a_{\mathbf{r}}^\dagger a_{\mathbf{r}} \Bigg].
\label{eq3}
\end{equation}
After Fourier transformation, the Hamiltonian becomes diagonal
in momentum space,
\begin{equation}
H_{\mathrm{SW}} = \sum_{\mathbf{k}} \Bigg[ 
\frac{4SA}{a^2} \left( 2 - \cos k_x - \cos k_y \right) + \frac{2SD}{a} \sin k_y + 2SK + M_{\mathrm{s}} B_x \Bigg] a_{\mathbf{k}}^\dagger a_{\mathbf{k}}.
\label{eq4}
\end{equation}
The corresponding magnon dispersion for $S=1$ is given by Eq.~\eqref{eq6} in the main text.

\subsection{Out-of-plane field}
\label{app:md_outplane}

We now consider an applied magnetic field perpendicular to the film plane. Starting from Eq.~\eqref{eq2}, we set $B_x=0$ and add the Zeeman term
\(
- M_{\mathrm{s}} B_z \sum_{\mathbf{r}} S^z_{\mathbf{r}}.
\)
In the presence of $B_z$, the equilibrium magnetization cants in the $xz$-plane by angle $\theta$, defined such that $\theta=0$ corresponds to ${\mathbf{M}}\parallel\hat x$ and $\theta=\pi/2$ to ${\mathbf{M}}\parallel\hat z$. Minimization of the uniform free energy yields
\begin{equation}
\sin\theta = -\frac{M_{\mathrm{s}} B_z}{2KS},
\qquad
\cos\theta = \sqrt{1-\sin^2\theta}.
\label{eq:theta}
\end{equation}

We rotate the spin basis around the $y$-axis by $\theta$,
\begin{equation}
\begin{pmatrix}
\tilde S^x \\ \tilde S^y \\ \tilde S^z
\end{pmatrix}
=
\begin{pmatrix}
\cos\theta & 0 & -\sin\theta\\
0 & 1 & 0\\
\sin\theta & 0 & \cos\theta
\end{pmatrix}
\begin{pmatrix}
S^x \\ S^y \\ S^z
\end{pmatrix},
\end{equation}
which leaves the exchange term invariant, since the rotation is uniform in space.

Applying the Holstein--Primakoff transformation in the rotated frame,
\begin{equation}
\tilde S_{\mathbf{r}}^+ = \sqrt{2S}\,a_{\mathbf{r}},
\qquad
\tilde S_{\mathbf{r}}^- = \sqrt{2S}\,a_{\mathbf{r}}^\dagger,
\qquad
\tilde S_{\mathbf{r}}^x = S - a_{\mathbf{r}}^\dagger a_{\mathbf{r}},
\end{equation}
and retaining only bilinear terms, we obtain the spin-wave Hamiltonian
\begin{equation}
\begin{aligned}[b]
H_{\mathrm{SW}} = \sum_{\mathbf{r}} \Bigg[
&-\frac{2SA}{a^2}\sum_{j\in\{x,y\}}
\big(a_{\mathbf{r}}^\dagger a_{{\mathbf{r}}+\hat e_j}
+ a_{{\mathbf{r}}+\hat e_j}^\dagger a_{\mathbf{r}}
-2a_{\mathbf{r}}^\dagger a_{\mathbf{r}}\big) + \frac{iSD}{a}\cos\theta\,
\big(a_{{\mathbf{r}}+\hat e_y}^\dagger a_{\mathbf{r}}
- a_{\mathbf{r}}^\dagger a_{{\mathbf{r}}+\hat e_y}\big)
\\
&+ \Big(2SK\cos^2\theta - SK\sin^2\theta - M_{\mathrm{s}} B_z \sin\theta\Big)\,
a_{\mathbf{r}}^\dagger a_{\mathbf{r}} -\frac{SK}{2}\sin^2\theta\,
\Big[(a_{\mathbf{r}}^\dagger)^2 + (a_{\mathbf{r}})^2\Big]
\Bigg].
\end{aligned}
\label{eq:Hsw_out_real}
\end{equation}

Fourier transforming yields
\begin{equation}
H_{\mathrm{SW}}
= \sum_{\mathbf{k}}
\left[
\epsilon_{\mathbf{k}}\,a_{\mathbf{k}}^\dagger a_{\mathbf{k}}
+ \frac{1}{2}\,\Delta_{\mathbf{k}}
\left(
a_{\mathbf{k}}^\dagger a_{- \mathbf{k}}^\dagger
+ a_{\mathbf{k}} a_{- \mathbf{k}}
\right)
\right],
\label{eq:Hsw_bog}
\end{equation}
with
\begin{align}
\epsilon_{\mathbf{k}}
&=
\frac{4SA}{a^2}(2-\cos k_x-\cos k_y)
+ \frac{2SD}{a}\cos\theta\,\sin k_y
+ 2SK\cos^2\theta
- SK\sin^2\theta
- M_{\mathrm{s}} B_z \sin\theta,
\\
\Delta_{\mathbf{k}}
&=
- SK\sin^2\theta.
\end{align}
To diagonalize the Hamiltonian in Eq.~\eqref{eq:Hsw_bog}, we perform a bosonic Bogoliubov transformation,
\begin{align}
b_{\mathbf k} &= u_{\mathbf k}\,a_{\mathbf k} + v_{\mathbf k}\,a_{-\mathbf k}^\dagger\,, \\
b_{-\mathbf k}^\dagger &= u^*_{\mathbf k}\,a_{-\mathbf k}^\dagger + v^*_{\mathbf k}\,a_{\mathbf k}\,,
\end{align}
with the normalization $|u_{\mathbf k}|^2 - |v_{\mathbf k}|^2 = 1,$ which ensures $[b_{\mathbf k},b_{\mathbf k}^\dagger]=1$. We parameterize $u_{\mathbf k} = \cosh\phi_{\mathbf k}$,  $v_{\mathbf k} = \sinh\phi_{\mathbf k}$, and choose the angle $\phi_{\mathbf k}$ so that the anomalous terms vanish, which yields
$\tanh(2\phi_{\mathbf k}) = \frac{\Delta_{\mathbf k}}{\epsilon_{\mathbf k}}$.
Under this transformation, the Hamiltonian becomes diagonal:
\begin{equation}
H_{\mathrm{SW}} = \sum_{\mathbf k} \omega_{\mathbf k}b_{\mathbf k}^\dagger b_{\mathbf k} + \mathrm{const},
\end{equation}
with the magnon dispersion given by Eq.~\eqref{eq:omega_outplane_main} of the main text for $S=1$.

\section{Equivalence of the power spectra of AABs and asymmetric bimerons}

\begin{figure}[!htb]
	\centering
	\includegraphics[width=0.85\linewidth]{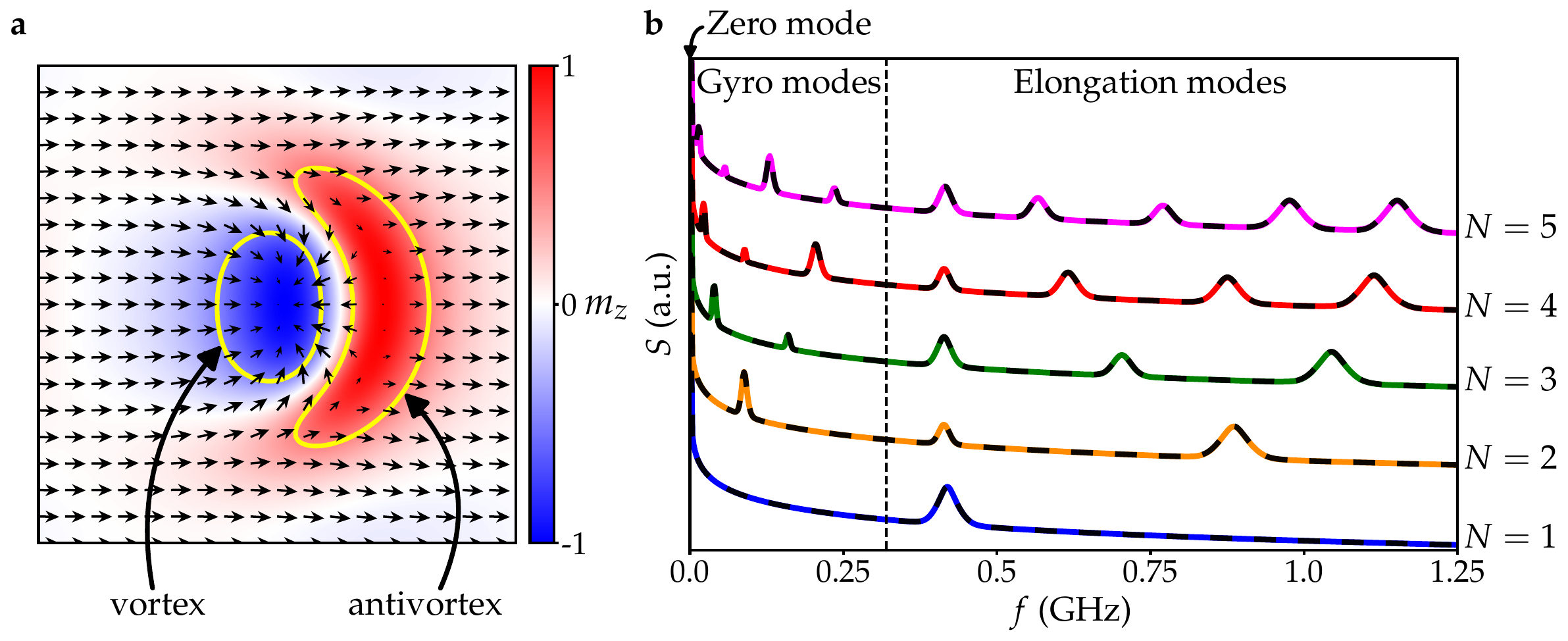}
	\caption{{\bf Comparison of the power spectra of AABs and asymmetric bimerons.} {\bf a} Magnetization configuration of a single asymmetric bimeron [cf. with Fig.~\ref{fig1}a of a single AAB in the main text]. Colors indicate the out-of-plane magnetization component. {\bf b} Colored solid curves show the spectra of clusters containing $N=$ 1 to 5 AABs, while black dashed overlays show the corresponding spectra of asymmetric bimeron clusters. The vertical dashed line separates the frequency ranges for gyrotropic and elongation modes. The power spectra $S(f)$ of AABs and asymmetric bimerons completely coincide for each $N$.}
	\label{fig1_s}
\end{figure}

\end{document}